\documentclass[amsmath,amssymb]{revtex4}

\usepackage{graphicx}
\usepackage{dcolumn}
\usepackage{bm}

\begin{document}

\title{Strongly anisotropic media: the THz perspectives of left-handed materials}

\author{Viktor A. Podolskiy$^1$, Leo Alekseev$^2$, and Evgenii E. Narimanov$^2$, 
\\ $^1$ Physics Department,  Oregon State University, Corvallis~OR~97331 
\\ $^2$ Electrical Engineering Department, Princeton University, Princeton NJ 08544 
\thanks{\vspace{6pt}\\ \centerline{VP:\tiny{vpodolsk@physics.orst.edu; http://www.physics.orst.edu/~vpodolsk }}
\\ \centerline{EN:\tiny{evgenii@princeton.edu; http://www.ee.princeton.edu/~evgenii}}
}} 

\begin{abstract}
We demonstrate that  non-magnetic ($\mu \equiv 1$) left-handed materials can be effectively used for waveguide imaging systems. We also propose a specific THz realization of the non-magnetic left-handed material based on homogeneous, naturally-occurring media. 
\bigskip
\end{abstract}

\maketitle

\section{Introduction}
The materials with negative refractive index \cite{veselago} (also known as left-handed media, LHM) have attracted a great deal of attention during recent years 
 \cite{pendry,pendryCyl,efros,kivshar,agranovich,smith,sridhar,trLine,shvetsPRB,pendryWires}.
  However, despite numerous efforts to bring LHMs to optical or even THz domain \cite{podolskiyWires,smithTHz,soukoulisTHz}, all modern realizations of these fascinating systems are limited to GHz waveguides\cite{smith,trLine,Parazzoli,PhysTodayLHM}.

LHMs associated with materials with simultaneously negative values of dielectric constant $\epsilon$ and magnetic permeability $\mu$, are typically based on several interconnected resonant structures. A fraction of these resonators is used to achieve negative dipole response, while others provide negative magnetic response \cite{smith,Parazzoli}. This basic LHM design however immediately leads to two problems. First, high-Q resonators needed in this approach, require extreme fabrication accuracy and uniformity across the the system -- something which is currently unachievable on a mass-production scale. 
Furthermore, the operation in a proximity of any resonance is typically 
accompanied by strong resonant absorption -- with the resulting loss of resolution
\cite{podolskiyResolut}. 

An alternative approach of using photonic crystals to achieve a negative phase velocity \cite{sridhar}, besides being sensitive to minute defects in the actual fabricated structure, typically yields direction-dispersive index of refraction with corresponding deterioration of optical properties. 

A new approach to obtain a left-handed response has been recently proposed in Ref.~\cite{podolskiyPlanar}. In contrast to the  resonant-based systems described above, the proposed material is non-magnetic (i.e. $\mu \equiv 1$), with the negative-$n$ response achieved in a waveguide configuration with anisotropic dielectric core. In this work we study the imaging properties of these non-magnetic LHMs, and propose their THz realization based on a homogeneous, naturally occurring material. 

\section{Wave propagation and imaging in non-magnetic LHMs} 

\begin{figure}
\centerline{\includegraphics[width=6cm]{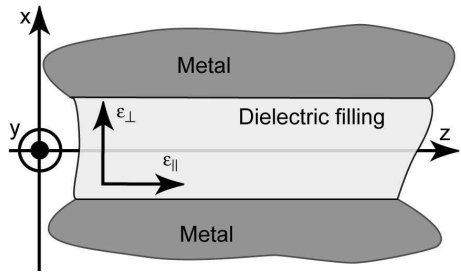}}
\caption{Schematic configuration of the non-magnetic left-handed waveguide} 
\label{figConfig}
\end{figure}

The proposed system is schematically shown in Fig.~\ref{figConfig}. It is represented by a planar (capacitor-type) waveguide with metal walls and anisotropic core. The dielectric constant of the core material is assumed to be uniaxial, with anisotropy axis perpendicular to the waveguide walls. 

An arbitrary electromagnetic wave propagating inside this system can be expressed in terms of its normal modes. Each of these modes is characterized by its polarization (with either electric [TE] or magnetic [TM] vector in the waveguide plane) and by its structure in $x$ (transverse) direction \cite{noTEM}. As it has been shown in Ref.~\cite{podolskiyPlanar}, the propagation of a mode in the proposed system is mathematically equivalent to propagation of a plane wave in an isotropic medium governed by the free-space-like dispersion relation
\begin{equation}
k_z^2+k_y^2=\epsilon\nu \frac{\omega^2}{c^2},
\end{equation}
where $k_z$ and $k_y$ are the propagation components of the wavevector of the mode, $\omega$ is its frequency, $\epsilon$ is equal to $\epsilon_{\perp}$ for TM modes and to $\epsilon_{||}$ for the TE ones, 
$\nu=1-\frac{c^2\kappa^2}{\epsilon_{||} \omega^2}$, and mode parameter $\kappa$ is defined solely by the mode structure in $x$ direction.

The effective refractive index of the waveguide system is given by $n^2=\epsilon\nu$. Thus, in order for the proposed planar system to support a propagating mode, the corresponding parameters $\epsilon$ and $\nu$ have to be of the same sign. The case of $\epsilon>0; \nu>0$ is typically realized in the planar systems with isotropic dielectric core \cite{landauECM} and corresponds to the 
``normal'' (``right-handed'') propagation. However, in contrast to this behavior, the case $\epsilon<0,\nu<0$ describes transparent structure with negative refractive index  \cite{podolskiyPlanar}. In such system -- similarly to the conventional ($\epsilon,\mu$) LHMs -- all phenomena directly related to phase velocity (e.g. Snell's law) are reversed. In particular, the planar slab of a non-magnetic LHM can be used to image objects within the waveguide. 

\begin{figure}[t]
\centerline{
\includegraphics[width=4cm]{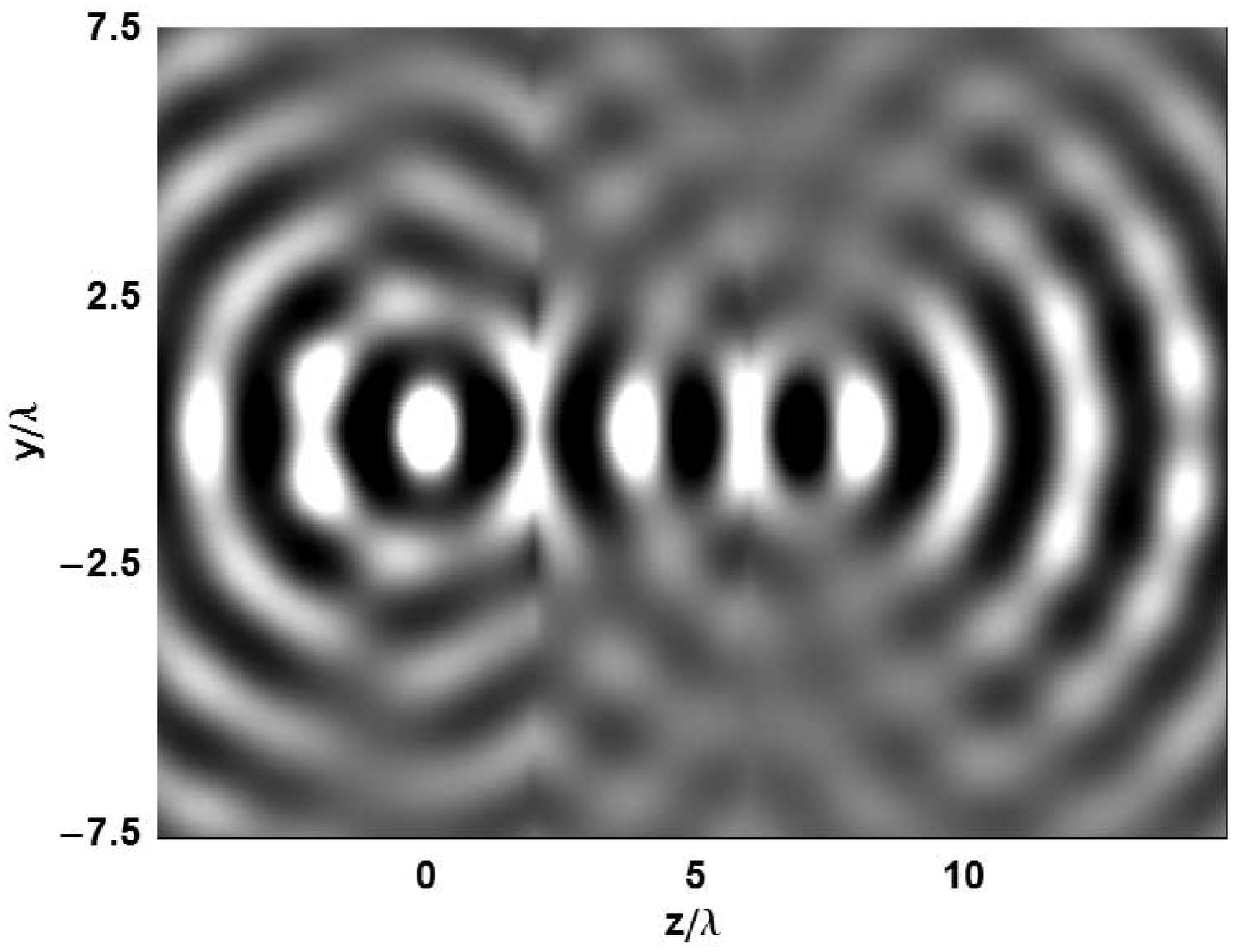}
\includegraphics[width=4cm]{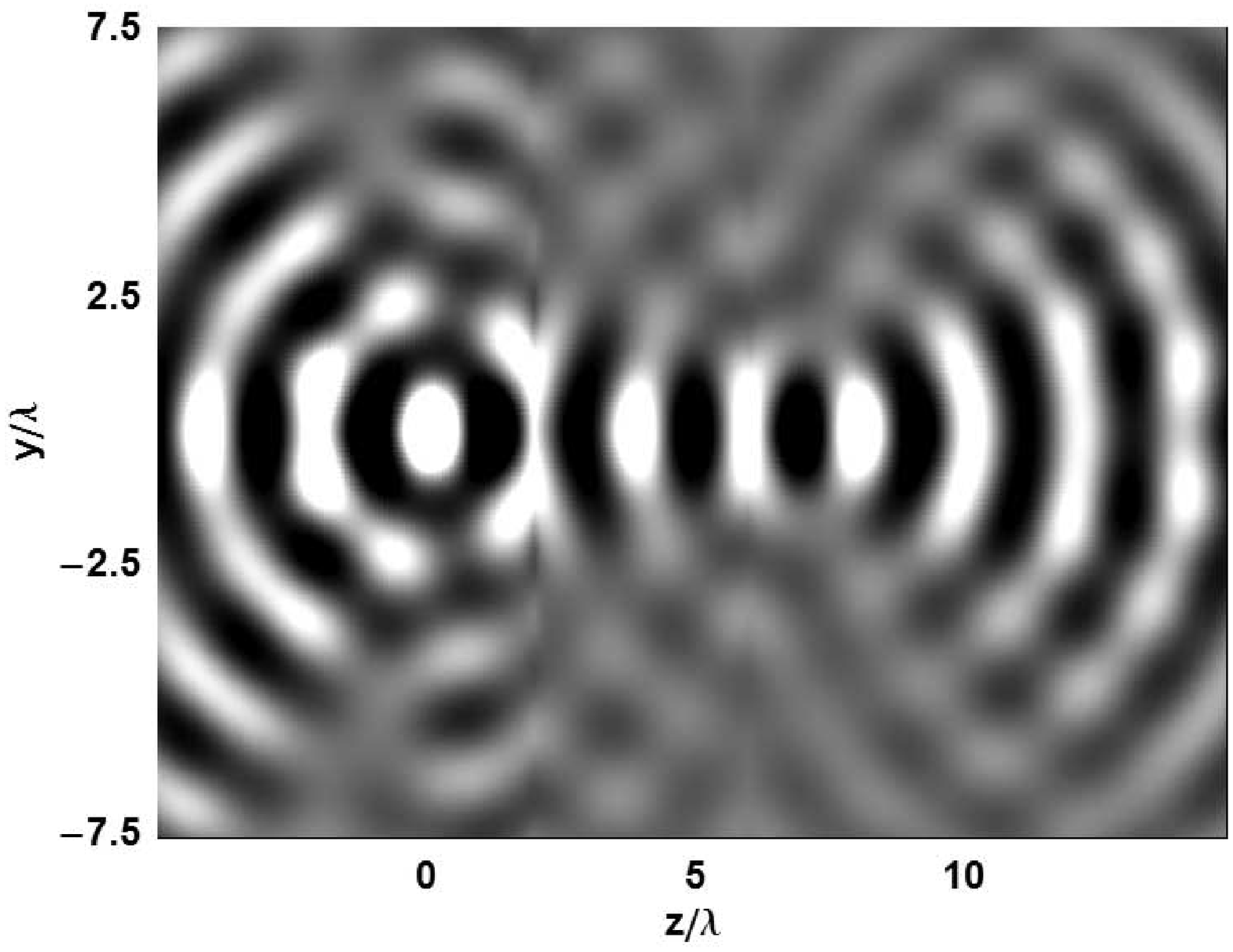}
\includegraphics[width=4cm]{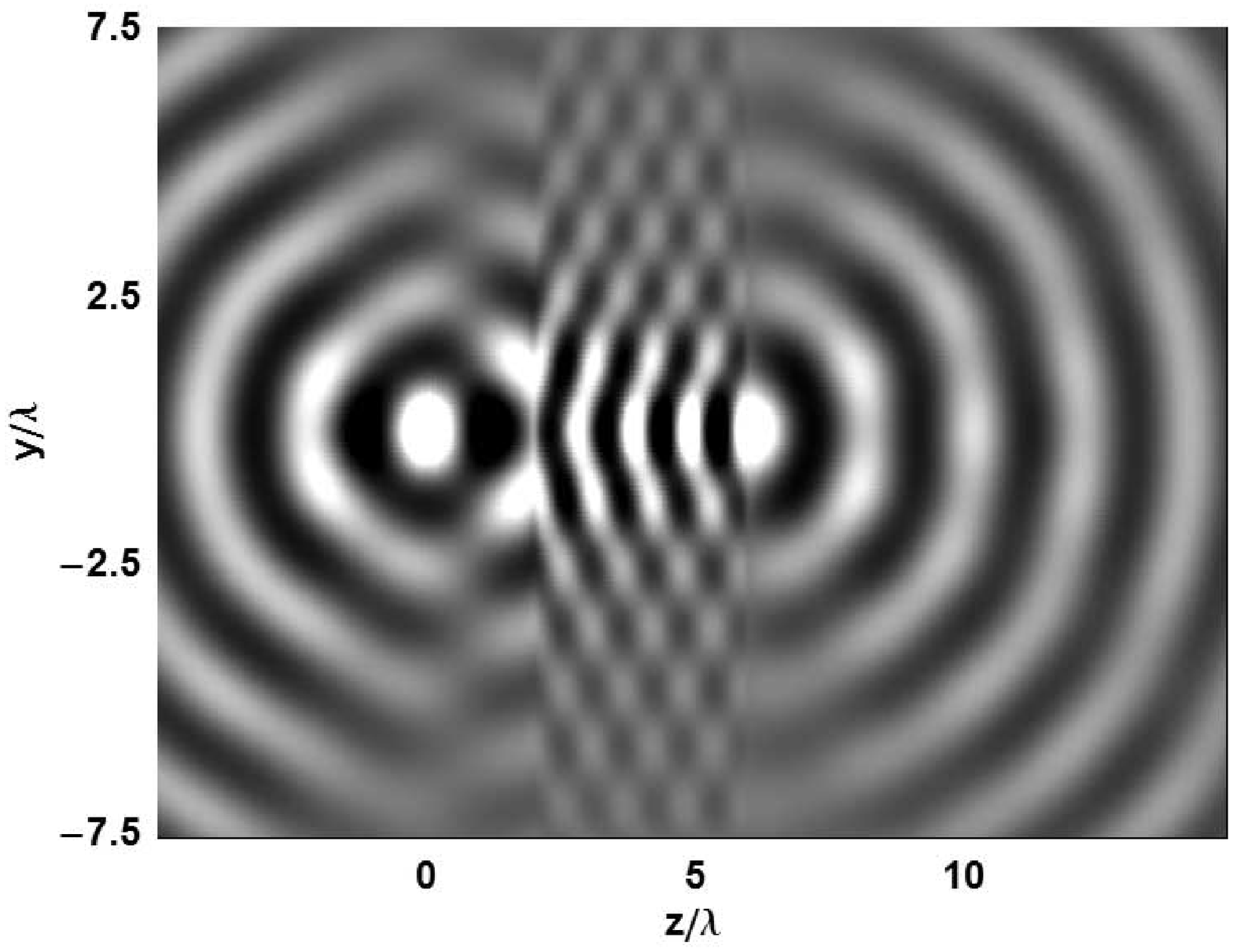}}
\caption{
Imaging using the proposed non-magnetic LHM system: the electric field component
$D_z$ is plotted; the left-handed material is extended from $z/\lambda=2$ to $z/\lambda=6$;  (left): imaging through the $\epsilon-\nu$ matched system [$\epsilon_{\rm RHM}=-\epsilon_{\rm LHM}=1/2; \nu_{\rm RHM}=-\nu_{\rm LHM}=1/2$]; (center): imaging through $n$-matched system $[\epsilon_{\rm RHM}=1/2; \epsilon_{\rm LHM}=-1/4; \nu_{\rm RHM}=1/2; \nu_{\rm LHM}=-1$]; (right) imaging through the non-matched left-handed system [$\epsilon_{\rm RHM}=1/2; \epsilon_{\rm LHM}=-1; \nu_{\rm RHM}=1/2; \nu_{\rm LHM}=-1$]} 
\label{figImaging1}
\end{figure}

\begin{figure}[b]
\centerline{
\includegraphics[width=4cm]{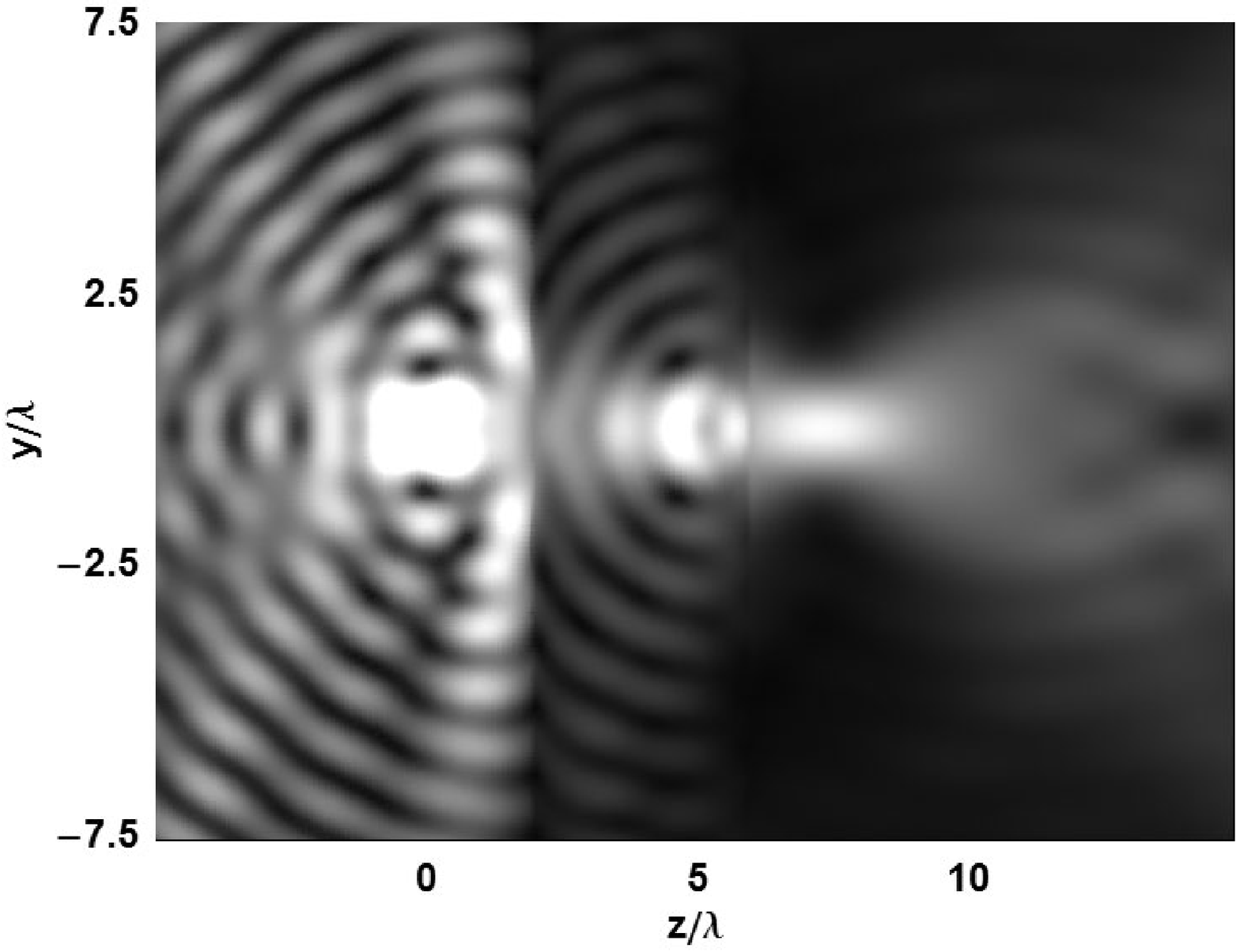}
\includegraphics[width=4cm]{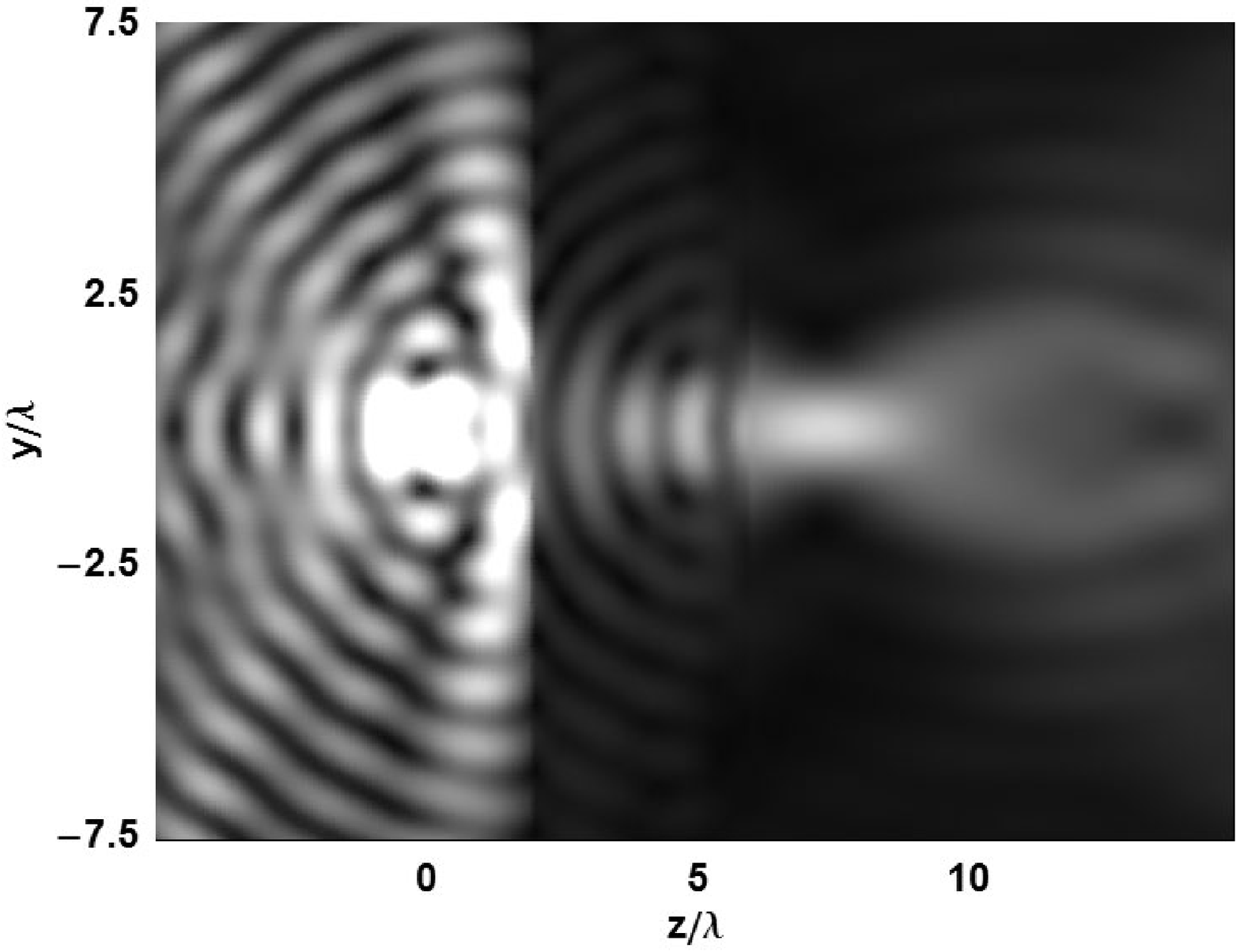}
\includegraphics[width=4cm]{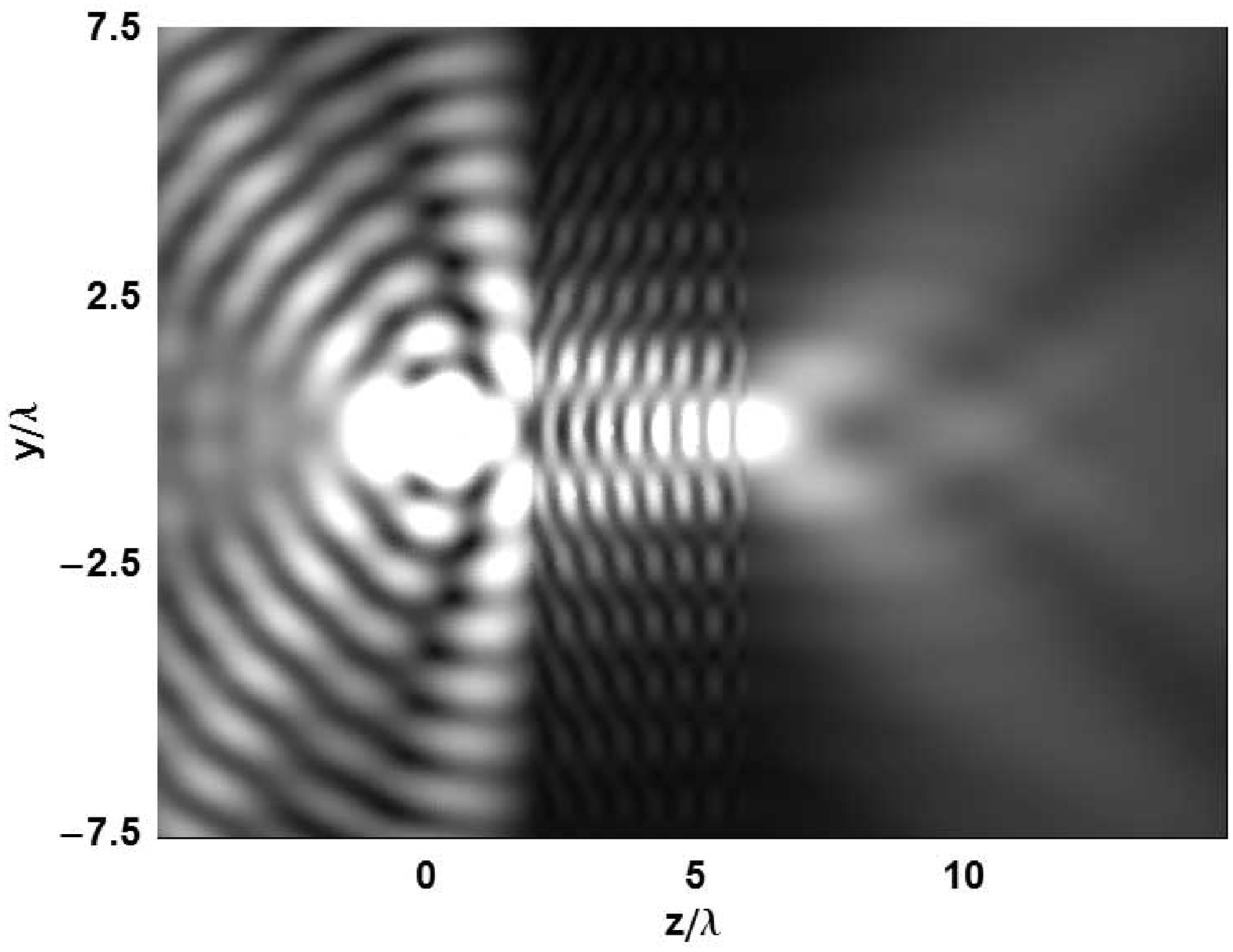}}
\caption{Same as Fig. \ref{figImaging1}, but showing the {\it total} intensity
$I \propto E_x^2 + E_y^2 + E_z^2$
} 
\label{figImaging2}
\end{figure}

Figs.~\ref{figImaging1}  and \ref{figImaging2} demonstrates such imaging process. To generate the data for these figures, we numerically solved Maxwell equations in the proposed system. In our numerical simulations we assumed a single-mode excitation with $\kappa=k/2$. Since the main focus of this paper is on THz perspectives of LHMs, we further assumed the waveguide walls to have infinite conductivity. 

In these simulations, a source of EM radiation was located at the origin (inside the RHM part of the structure), and its radiated wave was represented as a series of propagating TM modes with the same frequency and mode parameter $\kappa$. We then determined the transmission and reflection coefficients for each radiation component at front ($z=2\lambda$) and back ($z=6\lambda$) interfaces of the LHM insert, and computed the EM field throughout the system. 

The corresponding panels of Figs.~\ref{figImaging1} and \ref{figImaging2} represent  the same imaging processes, but show either the value of the $z$ component of the field (Fig.~\ref{figImaging1}) or the total intensity at a given point (Fig.~\ref{figImaging2}). The three different panels in each of these figures 
represent three principally different cases: (i) the case of complete $\epsilon-\nu$ match between LHM and RHM region [panel (a)], (ii) the case when RHM and LHM structures have the same refraction index, but the constitutive parameters are not matched [panel (b)], and (iii) the case when no parameters matched [panel (c)]. In each of the cases we can identify the formation of two images by an LHM slab -- the first image is formed inside the left-handed structure, while the second one appears in the right ``right-handed'' system. Note that the mismatch between the refractive index of right- and left-handed media in our case does not lead to a significant deterioration of the image quality. 

We also point out that if the source is positioned inside the LHM structure, one can effectively obtain two images on both sides of LHM region. This particular application may be useful for lasers based on sub-wavelength waveguides.

\section{Homogeneous non-magnetic LHMs}

As we discussed earlier, the realization of the non-magnetic left-handed materials requires a strongly anisotropic dielectric response. In this section we 
propose to use a material with effective electron mass anisotropy as one of the realizations of this non-magnetic LHM system. Note that while no natural material having simultaneously negative $\epsilon$ and $\mu$ have been found so far, the materials with strongly anisotropic effective mass do exist -- such as e.g. 
monocrystalline bismuth.

The high-frequency dielectric constant of a (semi-)conductor material containing the substantial amount of free electrons or holes is typically dominated by the dynamics of these free charge carriers. The resulting response is plasma-like, and with the dielectric constant being adequately described by the Drude model\cite{landauECM}: 
\begin{equation}
\label{eqDrude1}
\epsilon(\omega)=\epsilon_0+\frac{\omega_p^2}{\omega(\omega+i\Gamma)},
\end{equation}
where $\epsilon_0$ is the (frequency-independent) contribution of the bound electrons, $\Gamma$ describes inelastic processes, and the plasma frequency $\omega_p$ defined solely by free-charge-carrier concentration $N$ and effective mass $m_{\rm eff}$ 
\label{eqDrude2}
\begin{equation}
\omega_p^2=\frac{Ne^2}{m_{\rm eff}}
\end{equation} 

Two points can be immediately derived from the Eqs.~(\ref{eqDrude1}-\ref{eqDrude2}). First, the effective dielectric constant changes sign when the excitation frequency crosses the plasma frequency \cite{drude}. Second, the anisotropy of the effective mass immediately leads to the anisotropy of the plasma frequency, and correspondingly, to the anisotropy of the dielectric constant. It is therefore possible to achieve the desired strongly anisotropic dielectric constant provided that the material has strongly anisotropic effective carrier mass, and the operating frequency is between the plasma frequencies corresponding to the different effective mass directions. 

\begin{figure}
\centerline{\includegraphics[width=12cm]{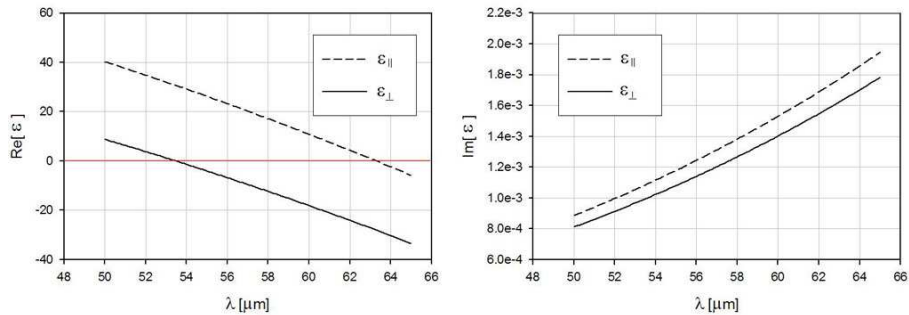}}
\caption{The real (a) and imaginary (b) parts of the dielectric function of bismuth at the liquid helium temperature, as functions of the wavelength in the THz frequency range. The ``perpendicular'' direction ($\perp$) is that of the $C_3$
crystallographic axis.  The red line at the left panel corresponds to the zero of the dielectric constant 
} 
\label{figBi}
\end{figure}

\begin{figure}
\centerline{\includegraphics[width= 8 cm]{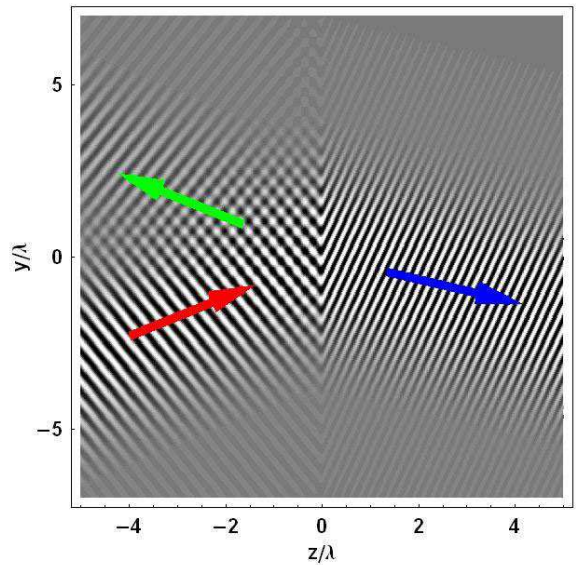}}
\caption{The refraction of a beam incident on the waveguide with bismuth core, for the wavelength within the negative index of refraction interval as described in the text, obtained from direct numerical solutions of the Maxwell equations. The arrows show the direction of the propagation of the incident (red), refracted (green) and reflected beams (blue). The medium on the left side of the boundary is an isotropic dielectric with $\epsilon = 55$; the medium on the right side is monocrystalline bismuth; the whole system is in a metallic waveguide with the thickness of $d = 4.5 \mu{}m$, and the (free space) wavelength is $\lambda = 61 \mu m$.} 
\label{figBiRef}
\end{figure}

We now illustrate the above considerations using the example of thin monocrystalline 
bismuth film, with the trigonal axis ($C_3$) perpendicular to the film surface. The
strong anisotropy of the effective masses of the electrons and halls in this semimetal (by a factor 
of up to $\sim 100$ for the electron pockets), leads to different values of the 
plasma frequency depending on the direction of the electric field. In particular, for 
our configuration the
experimental data of Ref. \cite{Boyle1960} yield  
$\omega_{p;\perp}=187 cm^{-1}$ and $\omega_{p;||}=158 cm^{-1}$. Therefore, for the frequencies $\omega_{p;||}<\omega<\omega_{p;\perp}$ (corresponding to THz domain) $\epsilon_{\perp}<0; \epsilon_{||}>0$ (see Fig.~\ref{figBi}). 

Another exciting property of left-handed system based on bismuth film is the extremely low material loss. In fact, the losses in Bi are so small that already in 1960s, bismuth mono-crystalline systems could yield the carrier mean free path
at liquid helium temperatures on the order of {\it millimeters} - see e.g. Refs. \cite{bismuth_mean_free_path}. As a result, the typical imaginary part of the dielectric constant of Bi can reach the values on the order of $10^{-6}$ (see Fig.~\ref{figBi}). Such a low loss is extremely advantageous for imaging and transmission applications \cite{podolskiyResolut}. 

Fig.~\ref{figBiRef} demonstrates the refraction of a wavepacket incident onto the Bi-based LHM system from a conventional waveguide . 

\section{Conclusions}
In conclusion, we have studied the imaging properties of the recently proposed non-magnetic negative-$n$ materials. We have demonstrated that these systems can be effectively used for imaging, and that refraction index or dielectric constant matching are not critical for the system performance. 

We have also proposed homogeneous, ultra low-loss, naturally occurring material (Bi film) with the strong anisotropy of the dielectric constant required for construction of non-magnetic LHMs, for THz frequencies.

\section{ Acknowledgments }

We acknowledge helpful discussions with R.~W.~Boyd, C.~Gmachl, J.~B.~Khurgin and M.~I.~Stockman.

This work was partially supported by NSF grants DMR-0134736, ECS-0400615, and Princeton Institute for the Science and Technology of Materials (PRISM).


\begin{thebibliography}{12}
\bibitem{noTEM} The TEM mode which may be present in a planar waveguide with perfectly conducting walls is neglected here since this mode (i)~is not present in LHMs considered here, and (ii)~does not couple to TE or TM modes.
\end{thebibliography}
\end{document}